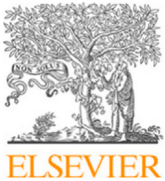
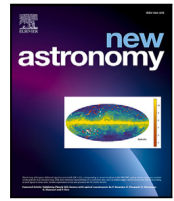

# A Search for recurrent novae among Far Eastern guest stars

Susanne M. Hoffmann [a,*], Nikolaus Vogt [b]

[a] *Michael-Stifel-Center and Institut für Praktische Informatik, Friedrich-Schiller-Universität Jena, Germany*
[b] *Instituto de Física y Astronomía, Universidad de Valparaíso, Chile*

ABSTRACT

According to recent theoretical studies, classical novae are expected to erupt every ∼$10^5$ years, while the recurrence time scale of modern recurrent novae ($N_r$) stars ranges from 10 to ∼100 years. To bridge this huge gap in our knowledge (three orders of magnitude in time scales), it appears attractive to consider historical data: In Far Eastern sources, we searched for brightening events at different epochs but similar positions that possibly refer to recurrent nova eruptions. Probing a sample of ∼185 Asian observations from ∼500 BCE to 1700 CE, we present a method to systematically filter possible events. The result are a few search fields with between 2 and 5 flare ups and typical cadences between $10^2$ and $10^3$ years. For most of our recurrence candidates, we found possible counterparts among known cataclysmic variables in the corresponding search areas. This work is based on an interdisciplinary approach, combining methods from digital humanities and computational astrophysics when applying our previously developed methods in searches for classical novae among Far Eastern guest stars. A first and rather preliminary comparison of (possible) historical and (well known) modern recurrent novae reveals first tentative hints on some of their properties, stimulating further studies in this direction.

## 1. Introduction

The eruption of a classical nova is a high energy event that can happen to an interacting binary star consisting of a white dwarf (WD) and a Roche-lobe filling component ('donor'). The permanent mass flow from the donor through the inner Lagrange point L1 causes an accumulation of hydrogen rich gas on the surface of the WD, leading to a thermonuclear runaway explosion with an increase in brightness of up to 19 mag and the ejection of a gas shell (Warner, 1995; Bode and Evans, 1989, 2008; Woudt and Ribeiro, 2014). Most novae are cataclysmic variables (CVs), with a late-type dwarf star as a donor but there are also 'symbiotic novae' (Munari, 2019) with giant stars as donors. Their amplitudes are smaller as compared to classical novae, because their pre-outburst brightness is determined by the red giant in the system and not by a faint accretion disk and/or the dwarf donor, as in classical novae. These is rather stable continuous burning on the WD's surface (Skopal et al., 2020) but apart from possible regular eruptions, these systems might also permit occasional classical novae. The fading after a classical nova eruption lasts many months or even years, and nova eruptions can occur many times in the CV with typical repetition rates between $10^4$ and $10^5$ years (Yaron et al., 2005). Therefore, for most novae only one event per star is observed.

However, there is a subclass called 'recurrent novae'[1] ($N_r$) with more than one eruption recorded. The Variable Star indeX (VSX) (Watson et al., 2006) returns 27 cataclysmic variables (CVs) that are known or suspected recurrent novae, ten of them outside our Galaxy (in M31 and LMC) and seven objects are not certain. The latest review on $N_r$ (Darnley, 2019) and the study by Schaefer (2010) on the photometric history and potential light curve templates of recurrent novae rely on the ten certain Galactic objects.

$N_r$ are grouped in three subclasses according to their orbital periods (Warner, 1995, 299–303), (Darnley, 2019): the T Pyx subclass has short orbital periods (hours) while the U Sco-type binaries have orbital periods of roughly one day. Their companion is evolved but still near the main sequence. The extraordinarily long $P_{orb} \geq 100$ d of the T CrB (or RS Oph) subclass implies that their companions are red giants (Fig. 1).

Fig. 1 shows the dependence of three important parameters ($t_3$ decay time, amplitude $A$ and recurrence frequency $t_r$) for nine of the ten Galactic $N_r$ cases (no $P_{orb}$ given for V2487 Oph). The T Pyx subclass is characterized by a relatively slow decay in the initial phase after eruption while the U Sco subclass shows a rapid decline (among the fastest known novae). The amplitudes in the T CrB subclass seem to be slightly smaller than those of the remaining $N_r$ cases, probably due to the larger brightness contribution of the red giant donor. Fig. 2 shows that neither $t_3$ nor $A$ are significantly correlated with the recurrence time $t_r$ but $A$ might vary a lot as the observed peak brightnesses occasionally vary independent of the subclass; cf. Fig. 3. With the current data, there is no obvious correlation but the statistics made from a sample of ten can provide only a first guess which stresses the wish to find more recurrent novae among historical observations.

---

* Corresponding author.
  *E-mail address:* susanne.hoffmann@uni-jena.de (S.M. Hoffmann).

[1] In the literature, there are the abbreviations RN and NR; the latter one is used in the VSX in particular. Here we use the standard symbol for novae (N) and indicate the recurrence by a subscript, similar to $N_a$, $N_b$, $N_c$, also written as NA, NB, NC.

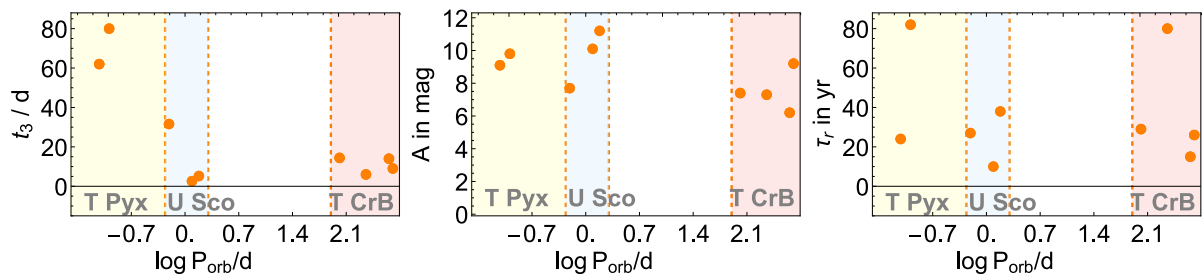

**Fig. 1.** Characteristics of the known recurrent novae in the Galaxy. Classified by their $P_{orb}$, there are three subclasses defined: the T Pyx-group (yellow), the U Sco-group (blue) and the T CrB-group (red). Their amplitude $A$ and recurrence time $\tau_r$ do not seem to depend on the class. Due to the $M_v \sim t_2$ relation for novae (della Valle, 1991, fig. 2), bright novae decline faster than fainter ones but the plot of the $t_3$-decline time does not show a relation to the class. ($A$, $t_3$ and $P_{orb}$ from Schaefer (2010, tab. 17), $\tau_r$ from Darnley (2019, tab. 1).

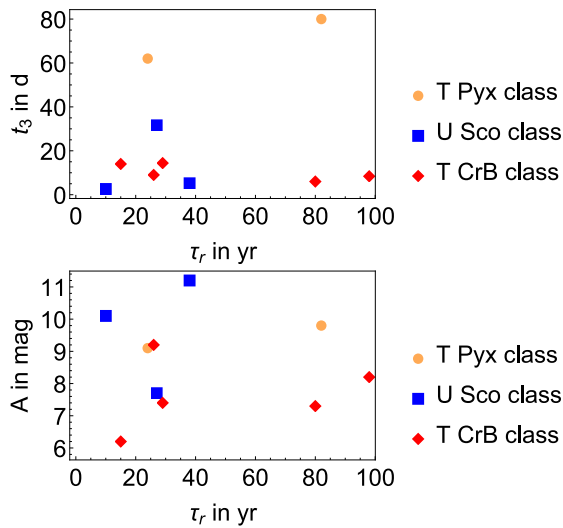

**Fig. 2.** The eruption amplitudes $A$ and decline times $t_3$ of known recurrent novae are not related to their recurrence times $\tau_r$ (data from Schaefer (2010, tab. 17) and Darnley (2019, tab. 1)).

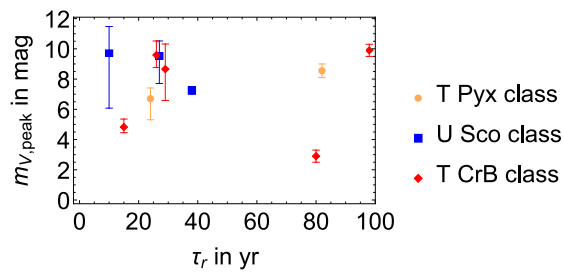

**Fig. 3.** Variability of peak brightnesses (data from Schaefer (2010, tab. 6-15) and VSX).

It is unknown, whether or not all cataclysmic variables (CVs) show this recurrent behaviour and on which time scales. Some models of the long term evolution of such systems were computed (Prialnik and Kovetz, 2005; Yaron et al., 2005), (Schaefer, 2010, fig. 10,11) and some consideration on observed behaviour of the past ~130 years are published e. g. in Vogt (1990), Leibowitz and Formiggini (2013),

Skopal et al. (2020).). In which way could it be possible to bridge the apparent gap in nova repetition time scales between observed modern $N_r$ cases (≤100 years) and classical novae (3 orders of magnitude larger, according to theory)? This raises the question of the existence of recurrent novae on millennia time scales. Here we suggest a method how to include even older observations.

## 2. Suggested historical recurrence phenomena

Among the uncertain $N_r$ is the suggestion of V529 Ori as recurrent nova (Ringwald and Naylor, 1997; Robertson et al., 2000): It is suggested to be identified with Hevelius' Nova Ori 1678 and also had been suggested as counterpart of a transient in 1894 by Packer (1894) and Shackleton and Fowler (1894). If it is connected to both, it might be a recurrent nova but the second transient is at all doubtful (Ashbrook, 1963, VSX,GCVS). Robertson et al. (2000) suggest a faint CV candidate ($V \sim 19$ mag) as possible counterpart that remains unconfirmed.

Another candidate for a recurrent nova is BZ Cam, a nova-like CV embedded in a multiple shell structure (Griffith et al., 1995; Bond and Miszalski, 2018), compatible with possible nova ejections by BZ Cam ~2, 5 and 8 millennia ago (Hoffmann and Vogt, 2020a). Due to the filaments in the nebula, it is certain that the CV erupted recurrently and this had been suggested already in the 1990s. However, with regard to the unknown expansion rates and the uncertainty of the proper motion, new measurements are required for an estimate of the recurrence timescale.

The classical nova from the symbiotic binary KT Eri, whose eruption in 2009 reached a visual maximum brightness of at least 5.4 mag in the regime of naked-eye visibility. It is also a candidate for a recurrent nova. Its relatively small eruption amplitude of ~10 mag is typical for symbiotic novae and comparable to that of modern recurrent novae. According to Hoffmann and Vogt (2020c), ancient historical records report a guest star event of 1431 near the same sky position, visible for 15 days and thus compatible with the modern classification of KT Eri as a rapid $N_a$-type nova. This opens the fascinating perspective of having perhaps identified a recurrent nova with a cycle of ~600 years.

Repetitions among a given positions within the 2.5 millennia of human observations could be a hint on recurrent novae on long time scales. Of course, there are also other possibilities to explain a particular repetition. It could be by chance that different objects occur at the same position and it could be biased by the purpose of divination that some asterisms are mentioned more often in chronicles than others. However, only the systematic search for recurrent novae could turn out some cases with the possibility of a real effect. Therefore, our study approaches this question with a first, relatively uncertain dataset that is typically accessed by astrophysicists when searching for possible historical counterparts, i. e. the sample of historical records from Far Eastern observers. This sample is silver standard (i. e. might include invalid cases such as comets that need to be checked due to a lack of further knowledge) and covers the recent ~2500 years.

## 3. Methods

For our earlier trials to identify historical records with known CVs (Hoffmann and Vogt, 2021), we used the relation of absolute magnitude and time of early decline: As generally novae with high amplitudes *A* decline faster than novae with small amplitudes, the duration of their visibility above a certain detection limit will be shorter. That means, the brighter a historical nova has been (relative to the CV's quiescence brightness), the shorter was the duration of its naked-eye visibility. The other way round, there were cases of given long durations of the visibility, leading to expected small amplitudes of the according novae (Hoffmann and Vogt, 2020b,a) and, thus, a brighter quiescence limit $m_{v,\mathrm{mod}}$ for the counterpart to be able to flare up to naked-eye visibility. The conditions of naked-eye peak brightness $m_{v,\mathrm{peak}}$ and the range of amplitudes defines the brightness filter for the quiescence counterpart: $m_{v,\mathrm{mod}} = m_{v,\mathrm{peak}} + A$. This method is also applicable in this study but with the constraint that peak brightnesses of the known recurrent novae vary typically between 0.5 and 2 mag while single peaks could deviate by even 4 mag from average (Fig. 3).

### 3.1. Starting point: Historical data used

The copies of records of transients in chronicles have been collected and compiled by scholars since the 19th century but usually, the nature of those sightings remained unclear. The simple information that something appeared does not allow any conclusion whether this was a star flaring up or a (tailless) comet whose motion is not reported. We deal with this uncertainty by not excluding anything in advance but perform an unbiased search for nova and supernova remnants at the given position (Hoffmann et al., 2020; Hoffmann and Vogt, 2020b,c). If none of the possible object categories is present, the record more likely refers to a comet.

If there is more than one object reported in 2.5 millennia at a given position, this could be chance coincidence or it could refer to a recurrent nova.

*Step 1:* As historical records are mostly preserved for divination, the preserved positions are given relative to asterisms instead of coordinates. In order to identify repetitions among them, we used a software routine to check all records that report stellar suspects and counted how often each asterism is mentioned: cf. Fig. 4 and the corresponding dates listed in Table 1.

*Step 2:* Afterwards, we compared the given positions: In some cases, only the asterism is given. In other cases, the position is further specified (like 'below [asterism]' or 'between stars *x* and *y* of [asterism]'). This allows to distinguish several position areas related to a particular asterism. The collections of historical records by Ho (1962), Hsi (1957) and Xu et al. (2000) have been filtered for records that (i) do not report a motion and (ii) do not report a tail. (iii) The certain supernovae of 1006, 1054, 1572, 1604 and the event in 185 are neglected. This list was used to count how often a particular asterism is mentioned.

*Step 3:* Some apparent repetitions in Table 1 might turn out as fake in the next step of the analysis when we compare the descriptions in more detail and define search fields for the remains of the historical transient. For instance, the events 396 and 667 are likely text corruptions, the three events in Kui refer to three different positions and for the event in Wei in 1203 it is unclear which of the three 'Wei' is meant: The given asterism is not visible at the given time and horizontal position, so the record is erroneous but there are several possibilities to correct the error.

Additionally, there is a bias (probably observational) that in years with comets there are highly likely also other transients reported; e. g. in 837 where comet Halley was great and afterwards, there were three guest stars. Hence, for some records at first glance it is unclear whether or not they still belong to the comet (as tailless) and a further study turns out that they are unlikely stellar transients (whatever else they are: part of a comet trail or mantic requirements). The most uncertain of these cases are excluded from our study.

*Limitation.* For all these reasons of uncertain interpretation, we limit this study to some of the events that are highlighted in the table. The scope of this study focuses on the development of a method how to deal with the question of potentially recurrent novae among Far Eastern transients. All results on particular objects are, of course, preliminary and need to be proven by further observations but the method demonstrates how to deal with the data.

### 3.2. Modern data used

The VSX was used for the search for cataclysmic variables (CVs), X-ray binaries (XB) and symbiotic stars and cross checked with the General Catalogue of Variable Stars (GCVS). CDS Simbad (Wenger et al., 2000) was used to find (potential) supernova remnants (SNRs), pulsars (PSRs) and potential nova shells (misclassified planetary nebulae, PNe) and for characterization of supernova remnants, the U Manitoba catalogue was referred to (Ferrand and Safi-Harb, 2012).

After filtering the huge output of these queries for the brightest (and thus most likely) objects at the areas of the given historical position, we additionally checked the individual light curves of all star candidates with the common services of light curve providers.

### 3.3. Ways of data mining

The method for the data mining for observational data in our search areas of various size had been developed by us earlier (Hoffmann et al., 2020).

Due to the novelty of this method, the fuzziness of the search fields (that are not sharp polygons or circles) and the reasonable scepticism to computer routines for filtering magnitudinal ranges that might be enrolled erroneously in the star catalogues, we performed the search for CV candidates twice:

*Alternative 1:* We plotted all CVs, symbiotic binaries and XBs together with the asterism lines into star charts. In the interactive mode, these star charts display the name of the star as tooltip by moving the mouse cursor over it. A human observer who had, thus, the same view on the asterisms as the ancient astronomer denoted the names of the CV candidates that can be described as given.

*Alternative 2:* The irregular area was approximated by circles. The coordinates and radii of these circles were used as input in the VSX query form (with groupings 'cataclysmic, symbiotic and X-ray binaries').

Both output lists were compared and we always found the same candidates. In both cases, the light curves of the candidates were cross-checked. Alternative 1 applies the brightness filter as computational routine. This routine is much more sophisticated than cutting the query in the VSX but it has to rely on correct catalogue data which is not always given. The advantage of this method is that it can find objects outside the (sharp) search field which cannot be ultimately defined in all cases (e. g. if asterisms overlap). In contrast, Alternative 2 finds only candidates within the defined circles but in case of incorrectly displayed magnitude ranges, this method could add stars that were displayed in our map but considered too faint. Furthermore, the first result must be mapped into star chart in order to ensure the positions of candidates still fit with regard to neighbouring asterisms.

As both alternatives return the same list of candidates, the method is again approved.

*Evaluating historical records:* In order to make the historical records useful, we developed a method of three steps that should always be performed:

1. Define the search field for modern counterparts in the sky and partition them into circles (coordinates with radii) for the VSX query (Hoffmann et al., 2020, Paper 3),

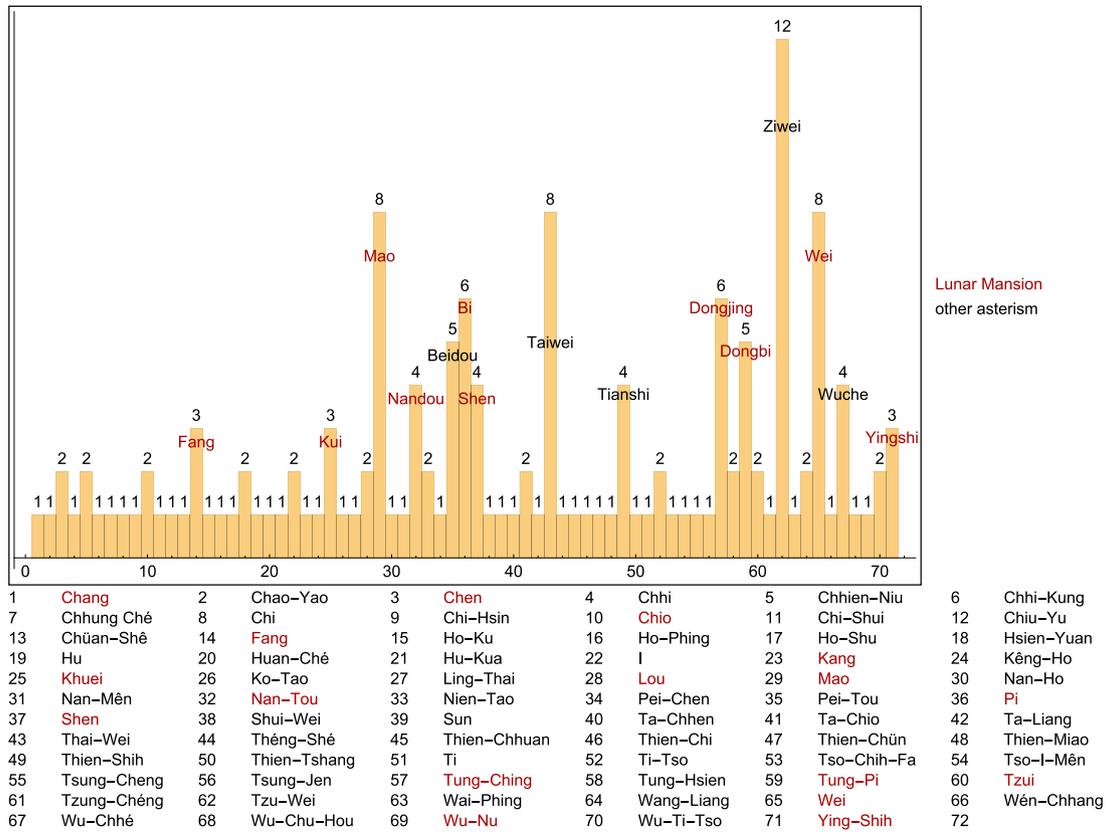

Fig. 4. Counting terms: Abundance of asterism names in ancient records of possible stellar transients. The list of asterism names in the legend is generated automatically from Ho (1962) and therefore uses the old spelling. The manually labelled bars use modern spelling of the same terms. Table 1 enlists the years per asterism. Lunar Mansion-asterisms (red) have special importance in the astrology and we therefore expect preservation of more observations in them than in other asterisms.

**Table 1**
The table entails the years of the records counted in Fig. 4 for the asterisms mentioned in more than two events. The second line gives an impression of the very different area sizes of these asterisms; the area estimates $A/°^2$ in square degrees are derived from our search fields covering them. Due to the merely phenomenological selection criteria this first selection is a silver standard; i.e. there could still be false positives that need to be filtered manually in a next step. Highlighted in blue are the events that we already discussed in Hoffmann et al. (2020) and the subsequent study. In cases where one year's record mentions two asterisms, the object appeared between them (as in case of 840 'between Dongi and Yingshi'); especially Mao and Bi are often mentioned together. Some of the records mentioned here will turn out as likely comets in the evaluation of the data. For the boxed asterisms all listed years are considered in this study; for the underlined asterisms, only some incidents are considered in this study.

|     | Ziwei | Taiwei | Wei | Mao | Bi | Dongjing | Dongbi | Beidou | Nandou | Shen | Tianshi | Wuche | Fang | Kui | Yingshi |
|-----|-------|--------|-----|-----|-----|----------|--------|--------|--------|------|---------|-------|------|-----|---------|
| $A/°^2$ | 2165 | 907 | 201 | | 201 | 254 | 201 | 263 | 101.5 | 531 | 2290 | – | 78 | | 308 |
| 1. | −76 | 64 | 393 | −46 | 304 | 107 | 840 | 158 | −47 | 483 | 123 | 667 | −133 | −75 | 840 |
| 2. | 85 | 126 | 708 | 396 | 396 | 278 | 877 | 305 | 386 | 773 | 149 | 668 | 436 | 1181 | 1113 |
| 3. | 269 | 222 | 1080 | 414 | 437 | 1073 | 329 | 837 | | 852 | 1461 | 683 | 1584 | 1297 | 1240 |
| 4. | 290 | 340 | 1203 | 449 | 837 | 1074 | 1123 | 1011 | | 956 | 1523 | 1596 | | | |
| 5. | 369 | 402 | 1224 | 639 | 639 | 1297 | 1388 | 1221 | 1415 | | | | | | |
| 6. | 537 | 419 | 1240 | 708 | 1313 | | | | | | | | | | |
| 7. | 541 | 617 | 1437 | 730 | 730 | 1596 | | | | | | | | | |
| 8. | 709 | 641 | 1600 | 877 | | | | | | | | | | | |
| 9. | 1175 | 1166 | | 1368 | 1368 | | | | | | | | | | |
| 10. | 1210 | | | | 1452 | | | | | | | | | | |
| 11. | 1373 | | | | | | | | | | | | | | |

2. *Hypothesis:* Query for cataclysmic variables (CVs) and symbiotic stars in these fields (Hoffmann and Vogt, 2020b, Paper 4)
3. *Antithesis:* check alternatives, i. e. supernovae remnants (SNRs), pulsars (PSRs) and other types of eruptive object (Hoffmann and Vogt, 2020c, Paper 5).
4. *Remaining option:* no stellar transient at all.

We successfully tested this method with ∼25 of 185 pre-selected events (criteria in Paper 3). It returned two historical records that are likely fake stars (e. i. corrupt text or a planet) and a list of 22 events with suggested counterparts of stellar transients and the remaining option of being comets.

## 4. Possibly recurrent events

The remaining events in Fig. 4 and Table 1 that report more than one sighting within a certain asterism could, in principle, refer to recurrent eruptions. However, as some search fields are some hundred square degrees, they contain hundreds of CVs and could also refer to eruptions from different stars: The more CVs are located in the search field, the higher the likelihood that more than one of them erupted within 2500 years. Thus, the likeliest recurrent novae could be identified in small search fields with low apparent object density (i. e. outside the clouds of the Milky Way).

In this section, we discuss the events in Dajiao, Fang and Yugui outside the Milky Way in more detail. Additionally, we suggest a strategy for events close to the Milky Way for the examples in Nandou and Wei.

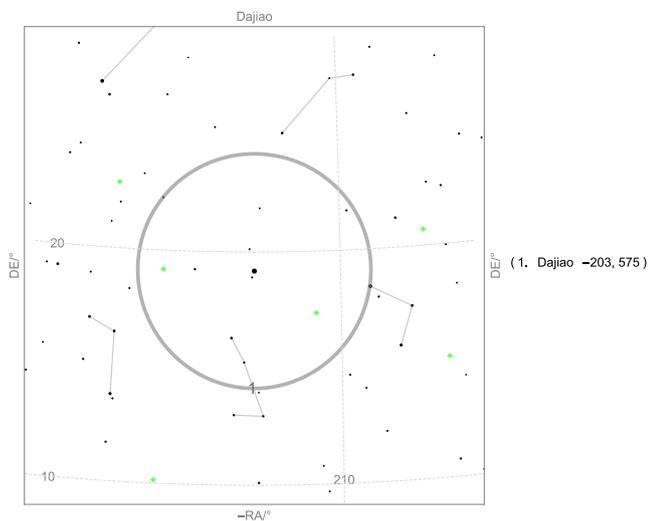

**Fig. 5.** Search field for Dajiao; green ◊-symbols mark the CVs. 'Dajiao' designates the single star in the centre of the map. The lines indicate neighbouring asterisms.

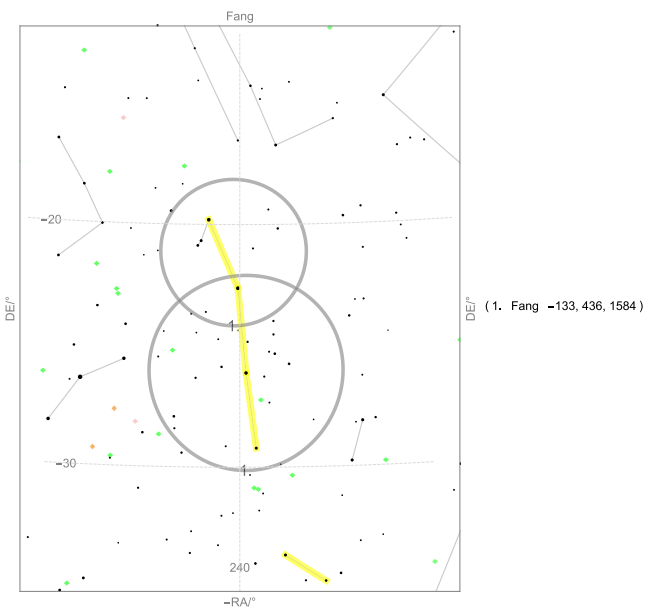

**Fig. 6.** Search field for Hipparchus' Nova; green ◊-symbols mark the CVs. It is connected with an apparition in the asterism Fang that consists of the stars connected with a yellow line in this map.

### 4.1. Search Field Dajiao 1 for events −203 and 575

The event in −203 had already been discussed by us: In Hoffmann and Vogt (2020c) we stated no SNR or PSR in this area but three bright CVs (see Fig. 5).

Herewith we add that actually two events are reported for this search field. The area contains two known novae, T Boo with a peak brightness of roughly 10 mag in 1860 which makes it undetectable for naked eye observers, and AB Boo that we already suggested as candidate for the event in −203 (Hoffmann and Vogt, 2020b). Its observed peak brightness in 1877 was 4.5 mag. Additionally, there are two relatively bright CVs in this search field: SDSS J143209.78+191403.5, a nova-like of VY Sct-type and the dwarf nova RX J1404.4+1723. If the events −203 and 575 refer to one and the same recurrent nova, this would mean that an event ~800 years later in the 14th century remained unobserved or the corresponding records were lost. In this case we expect a next eruption in roughly hundred years from now. However, the mean time interval could also be shorter, of the order of 700±100 years, and AB Boo could be the correct identification, referring to a third eruption observed in the 19th century.

Of course, it is also possible that both historical records refer to comets but there would be no remains. Thus, we first have to exclude the nova hypothesis before putting them into this category. An additional constraint of a recurrent nova will ease these efforts. In this case, the search circle is relatively small and the field is close to the Galactic pole which could make this study attractive.

### 4.2. The event −133 in Fang and possible recurrence in 436, 1584

Since Humboldt in the 19th century, the event in −133 is commonly treated as possible transient that inspired Hipparchus to make a star catalogue (Hoffmann, 2017, p. 8–10). The reason for this suggestion is a statement by Pliny the Elder that Hipparchus 'discovered a new star that was produced in his own age, and, by observing its motion on the day in which it shone, he was led to doubt whether it does not often happen that those stars have motion we suppose to be fixed.' Pliny and Bostock (1855, Book II, 24–26). Pliny's words shall mean that the object moved in the normal diurnal motion from the east to the west horizon. Neither the original observation by Hipparchus nor any other mentioning of this appearance are preserved from Greco-Roman Antiquity. Nothing is known on Hipparchus's biography; it is only known that he lived in the −2nd century because the Almagest uses some of his observations that are dated 265 years before Ptolemy's own ones dating to +137. Additionally, the coordinates in Hipparchus's (reconstructed) star catalogue fit the equinox of roughly −130, e. g. Grasshoff (1990, p. 30 and references therein). The only guest star in this century is the one in 134 CE (−133) but we expect many more: With three methods based on current star statistics, we estimated the statistical frequency of classical novae. We obtained the expectation of 1.4 to 10.7 novae per century with magnitude limit 2 and 4, respectively (Vogt et al., 2019; Hoffmann and Vogt, 2021). Thus, one to nine novae from this century have been unobserved or the records are lost.

Neither Pliny nor the Chinese chronicles preserve a duration or any further description of the phenomenon but if even Hipparchus realized it without expecting changes in the sky (or even believing that there are none), this transient in summer −133 should have been unmistakably bright and lasting a while. In the asterism Fang, there is no known supernova remnant but four old pulsars, the youngest one is PSR J1603-2531 with a characteristic age $\tau_c = 2.82 \cdot 10^6$ years. Thus, a supernova is highly unlikely.

Up to now, also no known nova remnant (shell) or suspect CV had been found (Hoffmann, 2017, p. 10) which made the whole scenario questionable. Yet, our statistics (Fig. 4) turns out that Fang is also mentioned in later records of transients: an 'anomalous star' that emerged on July 11th 1584 in Fang and the fuzzy[2] star on June 21st 436. In all cases, there is no further description. Might this refer to a recurrent nova?

If there was a classical nova in −133 and again in 436, the period of eruptions would be ~570 years. 570 years after 436, there was the impressing supernova of 1006 that outshone everything else for the chronicle and another ~570 (578) years later, there is a reported sighting in the small asterism of Fang. Thus, we are looking for a possible recurrent nova with a period of 570 to 580 years among the known the CVs and symbiotic binaries (see Fig. 6).

---

[2] 'fuzzy' or 'bushy' can also mean 'so bright that the eye sees rays' due to aberration (López-Gil et al., 2007; Protte and Hoffmann, 2020).

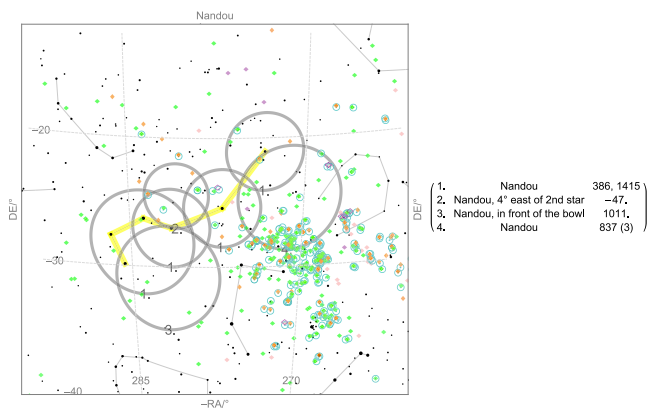

**Fig. 7.** Search fields for Nandou. The yellow line indicates the asterism line; green ◇-symbols mark the CVs.

Among the CVs in the field, only one target passes our filters: USNO-A2.0 0600-19894351, a CV of 17th magnitude without further sub-classification, possibly a nova-like binary because the total variability amplitude of only 1.5 mag given in the VSX is rather low for an ordinary dwarf nova. When we defined our brightness filter, we emphasized that CVs brighter than 18 mag are much more likely but systems down to 21 mag are possible (due to current knowledge). With this consideration, extending our CV search towards fainter magnitudes, a second candidate arises: Gaia20eoh, a dwarf nova of SS Cyg sub-type. Its orbital period is yet unknown but probably larger than 3 h, just within the range of those of most classical novae. The light curve of this dwarf nova in Gaia alerts reveals a mean quiescent magnitude of about $V = 19.5$ mag.

The remaining targets ASASSN-18pm, MASTER OT J155833.96-232213.9, SEKBO 106424.1762, Gaia17arr and ATLAS18oxn have very well-fitting positions but all of them are fainter than 20th magnitude and, therefore, less likely counterparts of nova events.

Extending our search also to the immediate vicinity of the borders of our search circle returns three additional targets: The eclipsing SU UMa-type dwarf nova V893 Sco (quiescent brightness of 15.5 mag) and another dwarf nova, ASASSN-15ib with 17.7 mag in quiescence but unknown orbital period. More interesting, however, could be the vicinity of T Sco, a classical nova with $V = 6.8$ mag at its maximum in 1860. With peak variability of 2 to 4 mag, possible naked-eye visibility is out of question. However, it is placed near the centre of the globular star cluster M80, resulting in only tiny chance to identify its quiescent counterpart in the extremely crowded background.

Summarizing, the relatively small search field slightly outside the Milky Way seems to provide recurrent eruptions of 570 to 580 years and there are several possible targets to study.

### 4.3. Search Fields for Nandou for events (−47,) (386), 837 c, 1011 and 1415

The Southern Dipper (Nandou) is mentioned in five records; the brackets in our headline indicate that we already treated these cases before. The handle of the Dipper touches the Milky Way and has, therefore, bright background but the bowl of the Dipper does not.

The descriptions of the transients are very different: The event dating in −47 is described as '4 degrees east of the 2nd star', a search field which is now – due to precession – north of the middle of the asterism line. The event 1011 happened 'in front of the bowl' (i. e. south of the asterism) and guest star 837 c was seen 'alongside Nandou and Tianyue' which is in the northwest of the asterism. Thus, the search fields for −47 (Nandou 2), 1011 (Nandou 3) and 837 c (Nandou 4) exclude each other (see Fig. 7).

However, there are two records that do not give the position more precise than 'in Nandou' and, thus, require to cover the whole area of the asterism with search circles. We already studied this area for the guest star in 386 CE that had been suggested as possible supernova (Clark and Stephenson, 1977) because it lasted three months. We suggested V1223 Sgr and V3890 Sgr as possible nova candidates in this field (Hoffmann and Vogt, 2020a, tab. 6).

V1223 Sgr is a relatively bright nova-like variable of DQ Her-type or a dwarf nova of Z Cam-type in permanent standstill. Either way, this CV has a high mass transfer and is, therefore, an ideal candidate for a recurrent nova. V3890 Sgr is a known recurrent nova with a cycle of 29 years (Darnley, 2019). Its peak brightness is known to be 7.1 mag leading to rather faint naked-eye visibility (if any). This way it would unlikely be recognized against the bright background of the Milky Way in this area. If it erupted stronger in 386 and/ or 1415, this would imply a rather large variability of $N_r$ peak brightnesses. Yet, a peak variability of 3 or 4 mag is still compatible with the observation; see Fig. 3. Thus, according to our current knowledge, V1223 Sgr is the more likely candidate for a recurrent nova on the time scale of centuries.

A recurrent nova could have been described differently in different epochs, i. e. it is possible that one astronomer described it 'in front of the bowl of Nandou' and another one says only 'in Nandou'. As these two search fields extend much beyond the other in a certain direction, the intersections of the fields Nandou 4 and Nandou 1 as well as Nandou 3 and Nandou 1 could also contain recurrent candidates.

Studying the intersection of fields Nandou 3 and Nandou 1 returns one of the options: If this refers to a recurrent nova, the sequence of observations would be 386, [unobserved ∼700], 1011, 1415 CE, i. e. a period of roughly 400 years. The search field contains the already suggested DQ Her-type V1223 Sgr.

Alternatively, the intersection of fields Nandou 4 and Nandou 1 would imply a possible recurrence of 450 – 580 years, namely in 386, 837 and 1415 CE. The search field touching the clouds of the Milky Way, returns 6 bright CVs and 1 Z And binary (V5759 Sgr = AS 270).

On closer inspection, the six potential CVs turn out to be uninteresting for this study: OGLE-BLG-DN-1017, OGLE-BLG-DN-091 and MACHO 161.24700.3300 are only bright at infrared wavelengths and seem to have no blue counterpart (cf. DSS) making them unlikely to be CVs. The dwarf nova OGLE-BLG-DN-0958 and the nova-like NSV 10530 are much too faint ($V \sim 18 - 20$ mag), and the potential DQ Her-type IGR J18173-2509 is in a field that is so crowded that a light curve cannot be generated bijectively. In this scenario, a good candidate could be the Z And-type star V5759 Sgr = AS 270. Symbiotic stars of the Z And class are binaries which are characterized by modest variations with optical amplitudes of 1–3 mag on timescales of weeks to years, e. g. AX Per (Leibowitz and Formiggini, 2013). Sometimes multiple rebrightenings are observed for the even longer time in the order of up to decades, e. g. Z And (Formiggini and Leibowitz, 1994; Skopal et al., 2000). In quiescence, symbiotic stars often show wave-like sinusoidal periodicities in accord with their orbital periods; examples are AG Peg, V1329 Cyg and V426 Sge (Skopal et al., 2020, their fig. 12).

In addition, some of the Z And stars show nova-like outbursts resulting from thermonuclear runaway as in the case of classical novae ('symbiotic novae' according to Munari (2019)). Due to the presence of the giant as the donor, typical amplitudes of eruptions in symbiotic novae are of 3–7 mag and, thus, much smaller than in classical novae, e. g. (Murset and Nussbaumer, 1994).

Still, there are also the symbiotic recurrent novae of T CrB subtype. In these cases, the novae on high mass WDs repeat on the human timescale, showing amplitudes of up to 11 mag (see Section 1). In the case of V5759 Sgr the WD mass is unknown. Fekel et al. (2007) give an orbital period of $671 \pm 7$ days (determined from the RV orbit) and the mass function $f(m) = 0.0189 \pm 0.0025$, implying possible masses of 1.5 and 0.5 solar masses for primary (M giant) and secondary (probably a WD), resp. However, since the orbital inclination is unknown, these are just lower limits, and we cannot exclude that V5759 Sgr could have a massive WD. This could be confirmed by its AAVSO light curve that reveals a stable quasi-periodic variation during the last $\sim$2000 days, with a period of about 330 days, just half of the orbital period. Since this is typical for the ellipsoidal shape of a nearly Roche-lobe filling binary component, this fact points towards the presence of a massive WD causing this deformation of the late-type giant companion. Therefore, V5759 Sgr must be considered as a valid candidate for a recurrent nova.

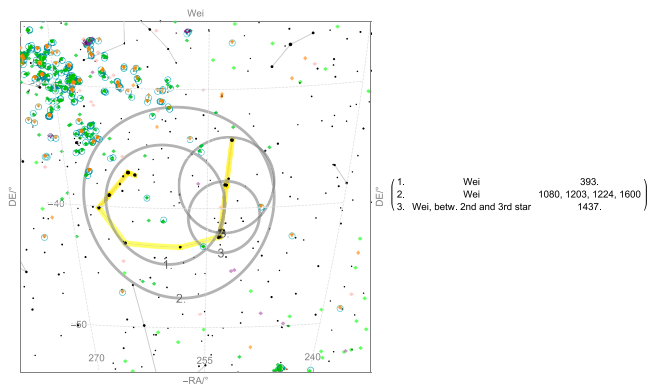

**Fig. 8.** Search field for Wei. The yellow line indicates the asterism line; green ◇-symbols mark the CVs.

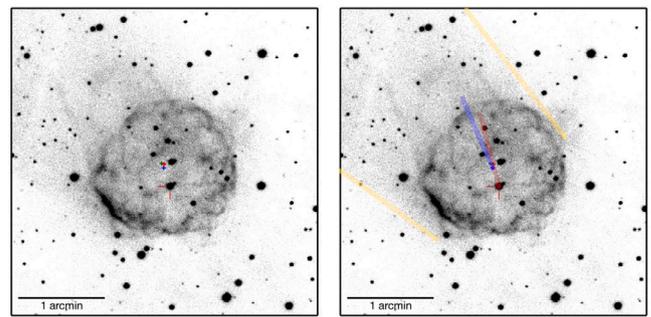

**Fig. 9.** Shara's Nova (CV+shell) in H$\alpha$ reproduced from (Shara et al., 2017b, fig. 1) (left) and with the 'tail' of the nebula (yellow) and the directions of proper motions (blue, red) marked by us (right).

Apart from CVs or symbiotic stars, in this search field we also found the high mass X-ray binary V4641 Sgr that possibly could flare up due to a tidal disruption event (TDE) at the black hole. Until now, there is no base for any prediction of this. The object has normally a V brightness between 9 and 14 mag, so a TDE-flare could brighten it to naked-eye visibility. The object lays in the field Nandou 4 and in the intersection of Nandou 1 and Nandou 4, so its consideration could play a role for the transients in 386, 837 and 1415 CE.

*4.4. Search field Wei 2 for events (393,) 1203, 1224, (1437) and 1600*

Like the Nandou asterism, Wei is also next to the Milky Way and the field has already been studied by F. R. Stephenson since the 1970s. Our search field Wei 2 extends over the whole asterism of Wei and, therefore, includes the (smaller) search fields of 393 and 1437 within it. The latter two events had already been considered by us: The search field of 1437 is between the second and the third star of the asterism and the search field of 393 is described as 'in the middle' which allows to neglect the edges — the question is only whether the middle of the area or the middle of the line is meant (see Fig. 8).

The huge search field Wei 2, close to the bulge of the Milky Way contains 123 CVs, among them 13 valid candidates for classical novae, more than 20 supernovae remnants and many pulsars. It appears brave to make any definite suggestion. All reported events could likely refer to six different objects flaring up. Therefore, we limit our analysis to a few better determined cases.

*Shara's nova, a proper motion-linked CV-nebula pair:* The event 1437 has been suggested to have a counterpart in Shara's age determined nova shell with proper motion-linked CV called GDS_J1701281-430612 (Shara et al., 2017b) but it does not fit the position between the second and third star in the standard identification: They suggest another counting of stars in the asterism and we asked whether the shell+CV-pair could refer to another transient. They assumed a systematic halving of the speed of the shell centre every 75 years due to interaction with the interstellar medium (Duerbeck, 1987). This factor certainly depends at least on the density if the medium. That means, in case of over- or underestimation of the reduction of shell speed, Shara's age of coincidence of the shell centre and the CV might be varied accordingly. That means, the epoch of the coincidence in the 15th century (or precisely in 1437) is not certain: The nebula and the CV could coincide also at another date in history. Could it refer to one of the reported guest stars in 1224 or 1600 although they do not fit the small error bars of the given position?

At first glance, Shara's shell does not show a structure of recurrent eruptions. This CV apparently does not produce recurrent novae on our considered time scale. However, the image they presented of this shell shows clearly a tail-like structure in the backward extension of the direction of motion (Fig. 9). It is not excluded that there are further structures as found in the 'tails' of BZ Cam and V341 Arae (Bond and Miszalski, 2018). We are looking forward to further investigations of this amazing CV-nebula pair.

*Possible recurrences:* A longterm recurrence is suspected in the given numbers of years: 393, [unobserved $\sim$800], 1203, 1600 could refer to a recurrence of roughly 400 years which would mean that there should have been another eruption in the past three decades. Another option would be a recurrence every $\sim$800 years if 1600 is a unique event and the sequence of years is only 393, 1203, [around 2000]. However, the event in 1203 is unreliable (Section 3.1, *Step 3*).

Re-interpreting the description 'in the middle of Wei' not as 'the area of Wei' but 'in the middle of the asterism line' of the Tail (Wei), the position of the event 393 could also refer to Shara's shell+CV and its vicinity. If the age determined by Shara et al. (2017b) is correct and the last big eruption happened in the 15th century, this could imply a recurrence every $\sim$1000 years. Conflicting could be the duration given in the records because in 1437 the event is reported for 14 days while the event in 393 lasted 7 months. Yet, in Hoffmann (2019, Sect. 3.4) we argued that the duration of 14 days could be a misunderstanding. In case something flared up around 400 CE and around 1400 CE, we do not expect a recent eruption.

Besides Shara's nova shell of likely but undetermined recurrence, it includes also three naked-eye novae with two of them having occurred around the year 2000: V992 Sco (1992, Na) and V1280 Sco (2007, Nb). The slow nova of V1280 Sco reached 3.8 mag and declined to naked eye invisibility within 2 weeks (see Fig. 10). That makes it unlikely to have produced a 7 months event in 393 CE. Additionally, it does not lay 'in the middle of Wei' as described in 393 CE but at the northern edge of our search field.

Due to its long visibility of 7 months, the event in 393 had been suggested to refer to a supernova (Clark and Stephenson, 1977). However, we emphasize that besides this possibility it could also refer to a slow nova of IGR J17195-4100 or V643 Sco (Hoffmann and Vogt, 2020a, tab. 6) but both of them are not known to have permitted a recent classical nova.

Summarizing, the idea of an event observed every 400 or 800 years appears unlikely on the base of the current knowledge.

The northeastern expansion of Shara's nova shell must be investigated in more detail for a final conclusion. The possibility of a recurrence on the time scale of a millennium remains.

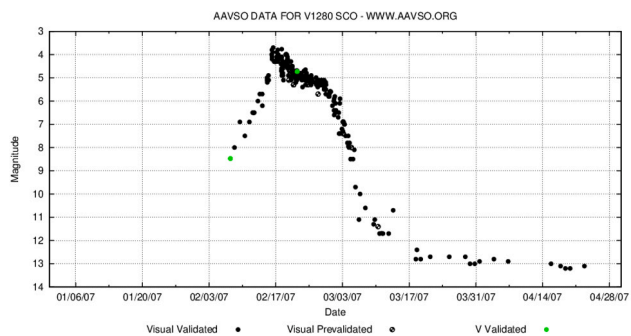

**Fig. 10.** Visual light curve of the Nb-type Nova Scorpii 2007 from V1280 Sco from VSX Light Curve Generator.

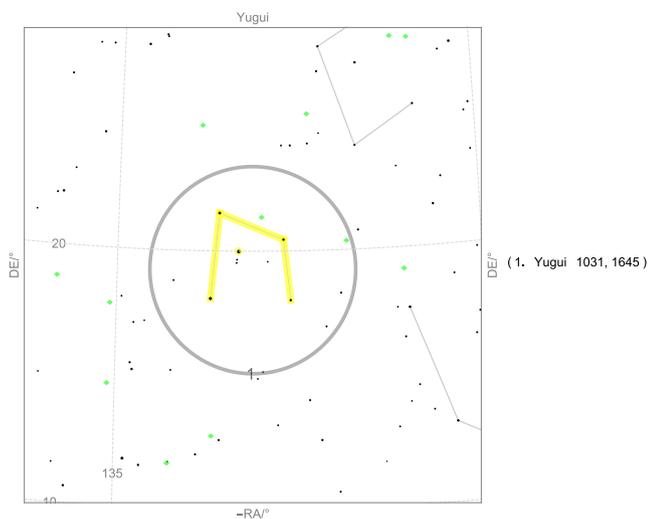

**Fig. 11.** Search field for Yugui. The yellow line indicates the asterism line; green ◇-symbols mark the CVs.

### 4.5. Search field Yugui for 1031 and 1645

For the sake of completeness, we mention this search field. There is no suspicious nebula in the field, no supernova remnant and no possibly misclassified planetary nebula. The search field in Cancer contains the old open cluster of Praeseppe (M44) but the area is rather dark, i. e. far from the Milky Way and from bright stars (see Fig. 11).

If this was a recurrent nova, a prior eruption in the 5th century could be unobserved and the next would be expected only in 200 years. However, we should find a CV that could reach naked eye visibility in this field. Shara et al. (2017a) already rejected the possibility that the old nova shell around AT Cnc refers to the event in 1645. CC Cnc is listed in the VSX as having 18.2 mag in minimum which is not quiescence but the observed minimum likely referring to a dwarf nova peak. CSS light curve shows that quiescence is fainter than 20 mag which exclude it as candidate. The only remaining object is the nearby DC white dwarf in the search field: HS 0819+2033 (Limoges et al., 2015). It is suggested as possible CV in the VSX and ASAS-SN Sky Patrol but the light curve is rather stable around $V \sim 16$ mag.

We again resume that there is no good candidate for a (recurrent) nova and relegate to the ambiguity of the term 'Yugui' that could refer to a very small asterism or to a whole lunar mansion. Instead of one recurrent nova, it is also possible that we deal with two unique events.

The two records from 1031 and 1645, both from Korea, report 'a large star entered Yugui' (Xu et al., 2000, p. 138 and 145). As they both use the verbum 'enter' and the adjective 'large', they could also refer to extended objects with a motion (comets) instead of stellar transients (Shara et al., 2017a; Hoffmann, 2019).

### 4.6. Likely impostors of recurrences

There are search fields that apply for more than one event but only by little chance contain a recurrently observed nova. They are discussed in the chronological order of the first year.

#### 4.6.1. Search Field Qianniu for events −4 and 588

This field was analysed earlier in our study without finding any candidate for a naked eye nova (or supernova). However, a new search revealed a possible candidate: The SU UMa-type dwarf nova SY Cap is listed with minimum $V = 19$ mag in the VSX and, thus, did primarily not pass our brightness filter. As the CSS light curve shows $V \sim 18$ mag, it is a valid candidate. A rather large nova amplitude (~14 mag) would be required for naked eye visibility.

If these two events refer to two eruptions from the same system and we could search for a recurrent nova on a timescale of ~600 years, additional eruptions around 1200 and 1800 would be expected. The 18th and 19th century were the great age of telescopic star mapping (new star catalogues by Flamsteed, Fortin, Bode and nebula catalogues by Messier and Herschel, cf. Latusseck and Hoffmann (2017)) and the start of big observational surveys culminating in the *Carte du Ciel* project. This suggests that potential novae of a naked eye brightness or slightly fainter would have been observed in this time.

As we cannot identify a bright CV candidate in a search field that covers only $41°^2$, we consider this suggested cyclical recurrence not impossible but unlikely.

#### 4.6.2. Search fields in Taiwei for events (64), 126, 222, 340, 419, 617 and (641)

The region of asterism Taiwei is rather large: The enclosure Taiwei (the Privy Council), covered by search field Taiwei 2, extends over more than 900 square degrees but the asterism is also named the 'Grand Tenuity' because the field is not very crowded.

Three events are reported in this area without giving more detailed locations: for the years 126, 340 and 419; they lay somewhere in the field Taiwei 2. However, in 64 CE and 222 CE the position is described close to a certain star ($\eta$ Vir, field Taiwei 1) close to the celestial equator, while for the years 617 (Field 3) and 641 (Field 4) the reports refer to positions 15 and 20° north, respectively. Field Taiwei 1, Field Taiwei 3 and Field Taiwei 4 exclude each other but all of them refer to one of the small asterisms within the Taiwei enclosure and, thus, each of them could possibly provide an event to be added to those of Field 2.

There are no SNRs or PSRs in either of the search fields.

A VSX search in the entire Field 2 reveals a total of 62 CVs. No CV was found in Field Taiwei 3, while in Field 1 the polar V379 Vir is the only possible identification that we already suggested in Hoffmann and Vogt (2020b) for the event +64.

**Field 1 and 2:** Assuming that Field 1 can be combined with the remaining three events in Field 2, a total of four nova eruptions every 60–100 years could have happened in the years 64, 126, 222, 340 and 419. For the event in 64 CE (and, thus, also 222), we found the faint and not convincing naked-eye candidate V379 Vir. Hence, this cadence is unlikely.

**Field 4 and 2:** Alternatively, Field Taiwei 4 contains three bright CV candidates, as already suggested in Paper 4 (tab. 2): the rather bright VY Sct-type nova-like star SDSS J122405.58+184102.7, the AM CVn type star IR Com and the poorly studied CV PG1119+149. If any of these identifications would be valid, the nova eruption cadence could be of the order of 200–300 years, for instance in a sequence 126, 419 and 641 CE.

Considering the remaining areas of Field 2, there are six additional, possible targets, among them four relatively bright dwarf novae whose mean magnitudes in their quiescent state are given here from light curves in Gaia alerts and/or CSS: RZ Leo (17.8 mag), TW Leo (16.0 mag), QZ Vir (16.0 mag) and V406 Vir (17.7 mag), and two

poorly studied other CVs: SDSS J123255.11+222209.4 (17–18 mag) and SDSS J124959.76+035726.6 (16.3 mag).

For all these possible eruption sequences the question arises why this activity is only observed within the relatively short time interval between the second and the fifths century, but neither earlier nor afterwards.

We do not dare to draw any conclusion on recurrence in this case but publish these thoughts in case later research brings up this possibility again. In this huge search field ($900°^2$), it is also well possible that the historical records refer to different phenomena.

### 4.6.3. Search field Beidou for events 158, 305, 329, 1123 and 1221

The event in 329 is exceptional because the text preserves that the object 'trespassed against' the Northern Dipper (see Papers 3, 4 and 5). That means, the object was close to the asterism line.

For the other four cases, it is not further specified where the transient appeared within the Big Dipper. There are several possible identifications among CVs: The well-known rather bright eclipsing nova-like star UX UMa (14.5 mag at minimum light), two other nova-like stars CT Boo and DW UMa and two rather faint polars, EV UMa and V496 UMa. Other possible targets are the dwarf novae IY UMa, V365 UMa or GP CVn.

There are ~150 years between the first two events (158 CE and 305 CE) and ~100 years between the second two (1123 and 1221). If they refer to a recurrent nova, there should be a CV that could have such a time scale for recurrence. Currently, none is known but the next decades of observational data will bring it up if it existed.

None of the 13 bright CVs in the field has a known nearby nebula that could be misclassified, so none of them is more likely than the others. In a search field that covers roughly $260°^2$, it is also well possible that the historical records refer to different eruptions.

### 4.6.4. Search field Zhen for events 247 and 275

In this $133°^2$ search field, there is no known SNR. Among the possible CVs in this field is CSS 100315:121925-190024, classified as a dwarf nova which is probably wrong because the CSS light curve shows a slow increase of its brightness from 19 mag to 17.5 mag in about 8 years, similar to VY Sct-type nova-like stars without any dwarf nova outburst activity. In addition, the area contains an unclassified CV (6dFGS g1222364-181050, 17.5 mag) and the dwarf nova SDSS J124602.02-202302.4 (17.5 mag in quiescence). TV Crv is a dwarf nova with super-outbursts whose quiescence is only a few tenths of a magnitude fainter than our 18.1 mag brightness filter: With a brightness normally between 18 and 18.5 mag, we do not exclude it as candidate.

Slightly outside the small asterism area but not yet in another asterism are the HD 106563, the undetermined cataclysmic variable EC 11560-2216 and the possible dwarf nova ASASSN-15fn (16.5 mag in peak).

Only 28 years had passed between the two ancient events in the same search field. If there is a recurrent nova with naked-eye visibility in this field at such a short eruption cadence, it would likely be known by us. One has to suppose large long-term variations on eruption amplitudes of long-term recurrent novae (see Section 5) to suggest this as recurrent nova. In contrast, two nova eruptions based on different CV targets could have occurred — especially with regard to the lack of any known supernova remnants and pulsars in this search field.

### 4.6.5. Search field Mao+Bi 5 for events 304 and 1452

The only remaining field of the five search fields in the area of Mao and Bi is the search field 'in Bi', around the old open star cluster of the Hyades ~$100°^2$ where no SNR is known. Three CVs are found in our search circle which had been defined rather huge. Two of the three CVs (MGAB-V247 and ASASSN-16pm) could likely also be described as 'between Mao and Bi' by a Chinese astronomer. No one of them is a good candidate. Remaining is the AM Her-type CSS 091109:035759+102943. Its normal brightness is between 17 and 18 mag (CSS light curve) which makes it a faint naked eye nova candidate and not highly likely in the Milky Way.

**Table 2**

Values of the recurrence cadences for ancient and modern recurrent novae with three or more recorded eruptions. Upper part: Ancient cases according to their asterisms and fields (see Table 4). Lower part: binaries of modern $N_r$. $N$ refers to the number of eruptions per star or field, $T \pm \Delta t$ to mean recurrence periods and errors suggested in Table 4 for ancient cases or given in Darnley (2019, tab. 1) for modern ones. A linear least square fit (cycle counts vs. eruption year) reveals the period value $\tau_r \pm \Delta\tau_r$, the standard deviation $\sigma$ and the ratio $\sigma/\tau_r$. (*) designates the uncertain case of V529 Ori that is suggested for 3 eruptions but in no case the identification is certain.

| Asterism/ | $N$ | Global estimate | Linear least squares fit | | |
|---|---|---|---|---|---|
| Star name | | $T \pm \Delta t$ | $\tau_r \pm \Delta\tau_r$ | $\sigma$ | $\sigma/\tau_r$ |
| | | y | y | | |
| Dajiao | 3 | $700 \pm 80$ | $687 \pm 31$ | 68 | 0.10 |
| Fang | 3 | $570 \pm 10$ | $573 \pm 1$ | 3 | 0.01 |
| Fang + T Sco | 3 | $285 \pm 10$ | $285 \pm 1$ | 5 | 0.01 |
| Nandou 1+4 | 3 | $515 \pm 65$ | $515 \pm 25$ | 36 | 0.07 |
| Nandou 1+3 | 3 | $205 \pm 10$ | $206 \pm 2$ | 6 | 0.03 |
| Wei 2+3 | 3 | $200 \pm 50$ | $188 \pm 14$ | 20 | 0.11 |
| Taiwei 1+2 | 5 | $80 \pm 20$ | $71 \pm 3$ | 12 | 0.17 |
| Taiwei 2+4 | 3 | $250 \pm 50$ | $258 \pm 20$ | 29 | 0.11 |
| Taiwei 2 | 3 | $95 \pm 15$ | $99 \pm 7$ | 15 | 0.15 |
| Beidou | 4 | $125 \pm 25$ | $118 \pm 2$ | 18 | 0.15 |
| T Pyx | 6 | $24 \pm 12$ | $23.5 \pm 2.5$ | 10.3 | 0.44 |
| RS Oph | 8 | $15 \pm 6$ | $11.9 \pm 0.3$ | 1.5 | 0.13 |
| V3890 Sgr | 3 | $29 \pm 1$ | $28.5 \pm 0.3$ | 0.4 | 0.01 |
| V745 Sco | 4 | $26 \pm 1$ | $25.7 \pm 0.2$ | 0.4 | 0.02 |
| CI Aql | 4 | $27 \pm 4$ | $27.6 \pm 1.3$ | 2.8 | 0.10 |
| U Sco | 14 | $10 \pm 1$ | $10.4 \pm 0.1$ | 1.2 | 0.12 |
| V529 Ori* | 3 | $215 \pm 5$ | $211 \pm 1$ | 3.55 | 0.02 |

### 4.6.6. Search field Yi 1 for events 421 and 561

The search area of the asterism Yi is covered by two large and two small circles with a total area of ~$357°^2$. It contains no SNR. Roughly seven CVs are found that are possibly bright enough to cause nova events with naked eye visibility: Two dwarf novae have rather bright quiescence magnitudes, TT Crt (15.3 mag) and ASASSN-15dw (15.1 mag), while the SU UMa-type star TU Crt (17.5 mag) is fainter. The quiescence magnitude of the dwarf nova NSV 5013 varies between $V = 17$–18 mag, according to the Gaia light curve.

Additionally, the SS Cyg-type dwarf nova ASASSN-15aa and the undetermined CV EC 11560-2216 are slightly outside the edge of the search field. The two nova-likes V0393 Hya and V0391 Hya lay in the southern extension of the asterism line in an area that does not belong to any asterism. With a lot of goodwill, these further four objects could also be considered for the given description of the position.

It is not excluded that one of the CVs mentioned here could have caused nova eruptions on a ~140 years recurrence time scale. If one of these objects erupted twice 1.5 millennia ago, the open question is why no further and recent brightening is reported on this target. Possible long-term variability on the nova recurrence cadence is totally unknown, and cases like this could give the first hints investigating these properties (see Section 5). Of course, a chance coincidence of two different nova events is also possible, and perhaps more likely in this case.

## 5. Conclusions

The results concerning possible novae among these recurrent events are visualized in Fig. 12 which combines the previously known information on recurrent novae (Schaefer, 2010, tab. 7), (Darnley, 2019, tab. 1) with the suggestions derived from our data (Table 4). They refer to the average time intervals $T$ between eruptions together with the maximal observed range of deviations $\Delta T$ including all supposed historical recurrent cases. The $T$ of modern novae show pronounced groupings around 10, 25 and 90 years (Fig. 2) but the last one is also populated by five possible ancient novae, see Fig. 12, left: green above blue columns. This histogram also shows a remarkable gap in the range $40 < T < 80$ y that matches the known qualities of $N_r$ as displayed in Fig. 2. The remaining ancient cases are rather homogeneously distributed over the range between 200 and 2000 years, implying a substantial extension of the total time span in our knowledge if there are some recurrent novae among them.

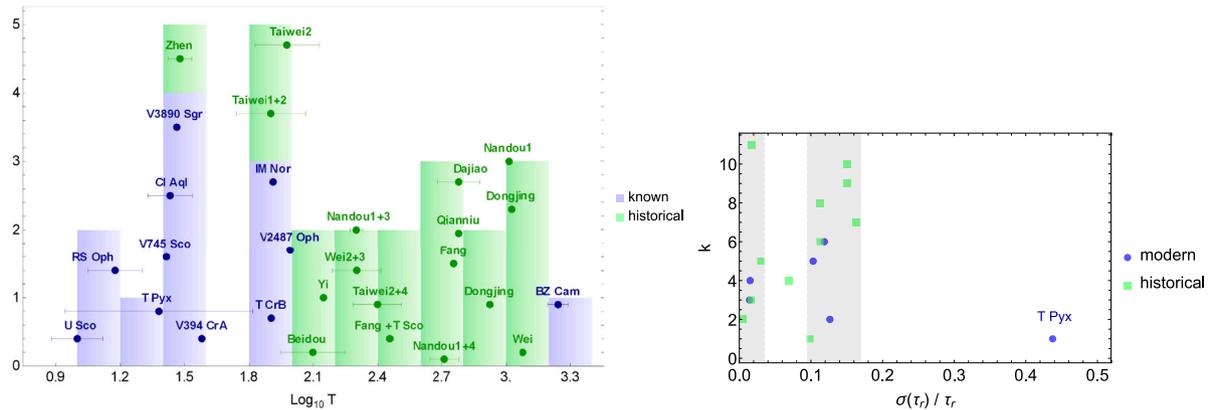

**Fig. 12.** Statistics of possible recurrent novae according to Tables 2 and 4, including historical and modern cases. Left part: histogram of the recurrence period $T$. The horizontal bars refer to the ranges given in Table 4. Right part: $\sigma/\tau_r$ vs. running index $k$. The grey areas mark the two groups mentioned in the text.

**Table 3**
The values variability of peak brightnesses $m_v$ is computed by us from the data in Schaefer (2010, tabs. 6-15) and the VSX (cf. Fig. 3). The huge range for U Sco comes from the well-covered small peak in 2020.

| | | | | | |
|---|---|---|---|---|---|
| $m_v$/mag | T Pyx $6.7^{+0.7}_{-1.4}$ | IM Nor $8.6^{+0.5}_{-0.4}$ | | | |
| $m_v$/mag | CI Aql $9.5^{+1.0}_{-1.8}$ | U Sco $9.7^{+0.5}_{-4.}$ | V394 CrA $7.25^{+0.25}_{-0.25}$ | | |
| $m_v$/mag | T CrB $2.9^{+0.4}_{-0.4}$ | RS Oph $4.8^{+0.5}_{-0.4}$ | V2487 Oph $9.9^{+0.4}_{-0.4}$ | V745 Sco $10.1^{+0.4}_{-0.4}$ | V3890 Sgr $8.7^{+1.7}_{-2.1}$ |

To estimate the mean time intervals between successive nova eruptions in a more quantitative way, we also applied linear least squares fits of the relation count number vs. year of eruption, whenever there are three or more eruption epochs available, and derived this way a recurrence period $\tau_r$, its error $\Delta\tau_r$ and the standard deviation $\sigma$. In most cases the eruption sequence presents a minimal time interval and sometimes small-number multiples of it, implying that missing eruptions were not observed or not reported. Only for the case in Beidou, which contains a total of four eruptions in two distant groups, we used the most probable guess of the number of cycles between these groups for the fit.

Table 2 contains the results of this analysis and gives a comparison of the global estimates, deduced from the data given in Table 4 with the parameters obtained by the linear least square fit. The eruption years of modern recurrent novae have been taken from Darnley (2019, tab. 1). Since $\Delta T$ comprises the entire observed range of deviations from a mean $T$ value while $\sigma$ is the standard deviation, we expect $\Delta T \sim 2.5\sigma$ in average, well compatible with the values in Table 2. It also lists the ratio $\sigma/\tau_r$ which is a measure of the 'regularity' of the eruption sequence: the smaller this ratio the tighter are nova eruptions following a periodic scheme.

The distribution of $\sigma/\tau_r$ shows agglomerations of our targets marked by grey zones in Fig. 12 (right): 5 cases with extremely low values $\sigma/\tau_r < 0.03$ and other 9 cases within the range $0.09 < \sigma/\tau_r < 0.17$. T Pyx presents an anomalously large $\sigma/\tau_r$ ratio due to its period increase from $\sim$11 to $\sim$36 years within one century. Historical and modern cases are present in both of the groups of small $\sigma/\tau_r$.

Could this be a first hint towards the possibility of a new classification system of recurrent novae? Shall they be classified according to their timescales $T$ of recurrence with regard to the apparent gap between 40 and 80 years? Or shall we classify them with their $\sigma/\tau_r$ ratio with regard to the two grey areas in Fig. 12 (right) and T Pyx as the prototype of a third group of changing recurrence time?

At the current state, we are only able to study these two basic parameters in a very preliminary way: the recurrence period $\tau_r$ and the eruption 'regularity' $\sigma/\tau_r$. Are they constant in time? Dramatic changes of the period $\tau_r$ are possible, as shown by T Pyx's strong increase by a factor 3 within one century. Independent information on past eruption sequences could arise from the analysis of multi-shell structures, as those in the nebula of BZ Cam (Bond and Miszalski, 2018). It looks as if their repetition varied, e. g. the filaments of the fourth and the fifth historical eruption (backwards in time from now) are closer to each other than the fifth and the sixth. Additionally, there had been higher activity of the system before the third (from now) eruption that is marked in black in Hoffmann and Vogt (2020a, fig. 3). A rough estimate of the kinematic age led to the suggestion of 1500 to 2000 years, compatible with the CV's proper motion. Still, this does not include the reduction of expansion velocity. Assuming a halving every 75 years (Shara et al., 2017b; Duerbeck, 1987), the speed would have reduced to the $10^{-6}$ or even $10^{-8}$ within the assumed $\tau_r$ which makes it difficult to draw any conclusion on these time scales without further measurements.

Is it possible that a relatively quiet CV or symbiotic Z And-type binary suddenly develops recurrent nova activity which is maintained for certain time and finally ceases?

This leads directly to the question of stability of peak magnitudes. In order to estimate their variability from modern recurrent novae, we have extracted this information from Schaefer (2010, tabs. 6-15) and the more recent data in the AAVSO database. Table 3 gives the mean peak magnitude values and the limits of their variations comparing all observed eruptions in each of the 10 established modern recurrent novae. In most cases these variations do not exceed $\pm 0.5$ mag but there are also differences $> 2$ mag. For the case of U Sco, the case with many observed eruptions, there is a total range of peak magnitudes of 6 mag.

**Table 4**
Catalogue of possible recurrences in historical records discussed in this work. The list is carefully handwritten after our analysis. The comments (last two columns) on the likelihood that the presented event(s) referred to novae at all and especially to recurrent novae are indicated by our estimate. The terminology scale is similar to the U Manitoba catalogue of supernova remnants http://snrcat.physics.umanitoba.ca/SNRtable.php: 'certain', 'likely', 'possible', 'suggested' and 'unlikely' (meaning: not impossible). Expected eruptions that lack observation are listed in [·], doubtful cases marked with '?' and those treated in earlier papers in blue.

| asterism | blue: already discussed Possible recurrences years | | | | | comment Nova? | recurrent? | $\Delta T_{estim.}$ y | search area | CV Candidat(s) name | type | reference |
|---|---|---|---|---|---|---|---|---|---|---|---|---|
| Dajiao | −203 | 575 | | | [1877] | likely | possible | 700 ± 80 | 78.5 | AB Boo | N: | Section 4.1 |
| Fang | −133 | 436 | [~1006] | 1584 | | likely | possible | 570 ± 10 | 78.5 | | | Section 4.2 |
| Fang+T Sco | −133 | 436 | [~1006] | 1584 | [1860] | likely | possible | 285 ± 10 | 78.5 | T Sco | NA, 1860 | Section 4.2 |
| Nandou 1 (only) | 386 | 1415 | | | | likely | possible | ~1030 | 101.5 | V1223 Sgr | DQ | Section 4.3 |
| Nandou 1 (only) | 386 | 1415 | | | | likely | possible | ~1030 | 101.5 | V3890 Sgr | NR: 30 y | Section 4.3 |
| Nandou 1+4 | 386 | 837 | 1415 | | | likely | possible | 450–580 | ~15 | V5759 Sgr | ZAND | Section 4.3 |
| Nandou 1+3 | 386 | [590,800] | 1011 | [1215] | 1415 | likely | possible | 200 ± 10 | ~17 | V1223 Sgr | DQ | Section 4.3 |
| Wei 1+2 | 393 | [~800] | 1203(?) | 1600 | [~2000] | suggested | unlikely | ~400* | 201 | V992 Sco | NA, 1992 | Section 4.4 |
| Wei 1+2 | 393 | [~800] | 1203(?) | 1600 | [~2000] | suggested | unlikely | ~400* | 201 | V1280 Sco | NB, 2007 | Section 4.4 |
| Wei 1+2 | 393 | | 1203(?) | | [~2000] | suggested | unlikely | ~800* | 201 | V992 Sco | NA, 1992 | Section 4.4 |
| Wei 1+2 | 393 | | 1203(?) | | [~2000] | suggested | unlikely | ~800* | 201 | V1280 Sco | NB, 2007 | Section 4.4 |
| Wei 1+2 | 393 | | | 1600 | | suggested | suggested | ~1200 | 201 | IGR J17195–4100 | DQ | Section 4.4 |
| Wei 1+2 | 393 | | | 1600 | | suggested | suggested | ~1200 | 201 | V643 Sgr | UGZ | Section 4.4 |
| Wei 2+3 | | 1224 | 1437 | 1600 | | possible | suggested | ~200 ± 50 | 78.5 | V1101 Sco | LMXB | Section 4.4 |
| Qianniu | −4 | 588 | | | | suggested | suggested | ~600 | 41 | | | Section 4.6.1 |
| Taiwei 2+1 | 64 | 126 | 222 | 340 | 419 | possible | suggested | 60–100 | 78.5 | V0379 Vir | AM | Section 4.6.2 |
| Taiwei 2+4 | | 126 | 419 | 641 | | possible | suggested | 200–300 | 57.3 | SDSS J122405.58+184102.7 | NL/ VY | Section 4.6.2 |
| Taiwei 2+4 | | 126 | 419 | 641 | | possible | suggested | 200–300 | 57.3 | IR Com | AM | Section 4.6.2 |
| Taiwei 2+4 | | 126 | 419 | 641 | | possible | suggested | 200–300 | 57.3 | PG1119+149 | CV | Section 4.6.2 |
| Taiwei 2 (only) | | 126 | [·] | [·] | 340 | 419 | suggested | suggested | 95 ± 15 | 908 | | | Section 4.6.2 |
| Beidou | 158 | 305 | 1123 | 1221 | | suggested | suggested | 100–150? | 263 | 13 bright CVs, no neb. | | Section 4.6.3 |
| Zhen | 247 | 275 | | | | suggested | suggested | 28? | 133 | | | Section 4.6.4 |
| Bi | 304 | 1452 | | | | possible | possible | 1150 | ~100 | CSS 091109:035759+102943 | AM | Section 4.6.5 |
| Yi | 421 | 561 | | | | likely | unlikely | ? | 357 | 8 bright CVs, no neb. | | Section 4.6.6 |

| Known novae with possible historical match | | | | | | | | | | | | |
| star name | type | years | | | | | | | | asterism | | |
| Shara's shell+CV | NA+UG/DQ+E | 393? | 1437? | | | certain | possible | ~1000 | | Wei 1+3 | | Shara et al. (2017b), Section 4.4 |
| BZ Cam | NL | 369? | | | | certain | certain | ~1500–2000 | 1614 | Zigong | | Paper 6 |
| KT Eri | NA+ZAND+E | 1431? | 2009 | | | certain | possible | ~600 | 274 | Jiuyou | | Paper 4 |
| V529 Ori | NR ? | 837(a) | 1678 | | | likely | possible | ~840 | | Dongjing | | |
| V529 Ori | NR ? | 837(a) | 1894 | | | likely | possible | ~1060 | | Dongjing | all identifications | |
| V529 Ori | NR ? | 837(a) | 1678 | 1894 | | likely | suggested | 211 ± 5 | | Dongjing | doubtful | |
| T CrB | NR | 1866 | 1946 | | | certain | certain | 80 | | Corona Borealis | | Darnley (2019) |
| V2487 Oph | NR | 1900 | 1998 | | | certain | certain | 98 | | Ophiuchus | | Darnley (2019) |
| V394 CrA | NR | 1949 | 1987 | | | certain | certain | 38 | | Corona Australis | | Darnley (2019) |
| IM Nor | NR | 1920 | 2002 | | | certain | certain | 82 | | Norma | | Darnley (2019) |

(*) In Wei, the year numbers suggest possible sequences of 400 or 800 years. Both cases rely on the historical record of 1203 that is highly questionable (see Section 3.1, *Step 3*) and in both cases we expect a nova eruption around the year 2000±10 y. As there is additionally no appropriate observation that might fit a naked-eye sequence, we consider this recurrence as 'unlikely' and neglect these suggestions in Table 4 but included only the suggestion of 1200 years.

**Table 5**
Search fields per event: The search field is approximated by circles covering the concerned field in the asterism. 'circNo' is the number of the circle within the field; the ♯-symbol means 'number of'.

| Field ID | in asterism | circNo | HIP | RA2000 | DE2000 | radius /° | comment | area/ °2 | ♯pairs | CVs | SNRs | PSRs |
|---|---|---|---|---|---|---|---|---|---|---|---|---|
| Field 1 | Beidou | 1 | 65802 | 202.324 | 53.1776 | 6 | | 263.057 | 0 | ~10 | 3 | 0 |
| | Beidou | 2 | 59882 | 184.209 | 56.2815 | 5 | | | | | | |
| | Beidou | 3 | 54495 | 167.24 | 59.2153 | 5 | | | | | | |
| | Beidou | 4 | 62956 | 193.507 | 55.9598 | 5 | | | | | | |
| | Beidou | 5 | 56510 | 173.77 | 54.7854 | 2 | | | | | | |
| Field 1 | Dajiao | | 69673 | 213.915 | 19.1824 | 5 | | 78.5398 | 0 | – | – | – |
| Field 1 | Dizuo | | 84345 | 258.662 | 14.3903 | 3 | | 28.2743 | 0 | 0 | 0 | 0 |
| Field 1 | Dongbi, betw.D. and Yingshi | | 116527 | 354.235 | 21.7231 | 9 | | 254.469 | 0 | – | – | – |
| Field 2 | Dongbi, west of | 1 | 117718 | 358.122 | 19.1203 | 6.5 | | 245.83 | 0 | 1 | 0 | 2 |
| | Dongbi, west of | 2 | 117332 | 356.872 | 25.5795 | 6 | | | | | | |
| Field 3 | Dongbi, south of | | 1038 | 3.2503 | 12.331 | 3 | | 28.2743 | 0 | 2 | 0 | 0 |
| Field 4 | Dongbi | | 578 | 1.75834 | 22.8445 | 8 | | 201.062 | 0 | 2 | 0 | 3 |
| Field 1 | Dongjing | | 32451 | 101.588 | 18.8377 | 9 | | 254.469 | 0 | 14 | 4 | 11 |
| Field 2 | Hu star, Dongjing (LM) | | 32007 | 100.309 | −40.3498 | 5 | | 78.5398 | 0 | 7 | 0 | 1 |
| Field 3 | Dongjing, below | | 29426 | 92.985 | 14.2088 | 6 | | 113.097 | 0 | 8 | 3 | 11 |
| Field 1 | Duanmen, near Pingxing | | 58510 | 179.987 | 3.65521 | 3 | | 28.2743 | 0 | 0 | 0 | 0 |
| Field 1 | Fang | 1 | 78265 | 239.713 | −26.1141 | 4 | | 78.5398 | 0 | 1 | 0 | 4 |
| | Fang | 2 | 78467 | 240.274 | −21.1557 | 3 | | | | | | |
| Field 1 | Kui, between K and Lou | | 7390 | 23.8266 | 23.0576 | 5 | | 78.5398 | 0 | 1 | 0 | 0 |
| Field 2 | Chuanshe (LM Kui) | | 7078 | 22.8073 | 70.2646 | 4 | | 50.2655 | 0 | 1 | 0 | 4 |
| Field 3 | Kui | 1 | 4095 | 13.1444 | 34.6394 | 7 | | 307.876 | 0 | ~3 | – | 1 |
| | Kui | 2 | 4761 | 15.2948 | 26.473 | 7 | | | | | | |
| Field 1 | 5° east of Juanshe (LM Mao) | 1 | 20252 | 65.1027 | 34.5667 | 5 | | 161.259 | 1 | 3 | 0 | 2 |
| | 5° east of Juanshe (LM Mao) | 2 | 20395 | 65.501 | 39.934 | 5 | | | | | | |
| | 5° east of Juanshe (LM Mao) | 3 | 19428 | 62.4119 | 31.6521 | 3 | | | | | | |
| Field 2 | Mao,south of | 1 | 18116 | 58.0962 | 18.5977 | 5 | | 157.08 | 0 | 0 | 0 | 2 |
| | Mao,south of | 2 | 16813 | 54.1014 | 20.0676 | 5 | | | | | | |
| Field 3 | Mao+Wei[Ari] | | 15696 | 50.5496 | 27.6076 | 5 | | 78.5398 | 0 | 1 | 0 | 0 |
| Field 4 | Mao, north of | 1 | 17547 | 56.3648 | 28.6687 | 3.5 | | 88.75 | 1 | 3 | 0 | 0 |
| | Mao, north of | 2 | 18972 | 60.9763 | 28.126 | 4 | | | | | | |
| Field 5 | Mao+Bi (Bi) | | 19877 | 63.9429 | 15.4007 | 8 | upper half circle | 201.062 | 0 | 3 | 0 | 1 |
| Field 1 | Nandou | 1 | 92927 | 283.995 | −28.1302 | 4 | | 101.451 | | | | |
| | Nandou | 2 | 90496 | 276.993 | −25.4217 | 1.5 | | | | | | |
| | Nandou | 3 | 92041 | 281.414 | −26.9908 | 3 | | | | | | |
| | Nandou | 4 | 89341 | 273.441 | −21.0588 | 3 | | | | | | |
| Field 2 | Nandou, 2nd star | | 91858 | 280.918 | −24.5049 | 2.5 | | 19.635 | | | | |
| Field 3 | Nandou, in front of the bowl | | 92184 | 281.824 | −30.8617 | 4 | | 50.2655 | 0 | 4 | 1 | 1 |
| Field 4 | Nandou | | 88560 | 271.242 | −24.6808 | 4 | bulge inside | 50.2655 | 4 | | | |
| Field 1 | Niandao, southeast of N. | | 98767 | 300.906 | 29.8968 | 6 | | 113.097 | 1 | > 17 | 18 | > 43 |
| Field 1 | Qianniu | 1 | 100345 | 305.253 | −14.7815 | 3 | | 41.0034 | 0 | 3 | 0 | 2 |
| | Qianniu | 2 | 99572 | 303.108 | −12.6175 | 1.5 | | | | | | |
| | Qianniu | 3 | 100881 | 306.83 | −18.2117 | 2 | | | | | | |
| Field 1 | Shen, east of | 1 | 28981 | 91.7344 | −3.34118 | 7 | | 204.204 | 1 | 1 | 2 | 7 |
| | Shen, east of | 2 | 28542 | 90.3802 | 2.90926 | 4 | | | | | | |
| Field 2 | Shen | | 26870 | 85.5734 | 2.36718 | 13 | | 530.929 | 1 | 15 | 2 | 5 |
| Field 1 | Taiwei, inside Zuoyemen | | 60129 | 184.976 | −0.666833 | 5 | | 78.5398 | 0 | 2 | 0 | 0 |
| Field 2 | Taiwei | | 59255 | 182.321 | 11.2917 | 17 | | 907.92 | 1 | 18 | 0 | 6 |
| Field 3 | Taiwei, in Wudizuo | | 57632 | 177.265 | 14.5721 | 4 | | 50.2655 | 0 | 0 | 0 | 1 |
| Field 4 | Taiwei, trespassed against Langwei | 1 | 59078 | 181.732 | 20.4929 | 1.5 | southern half | 57.3341 | 0 | 2 | 0 | – |
| | Taiwei, trespassed against Langwei | 2 | 60957 | 187.43 | 20.8961 | 4 | southern half | | | | | |
| Field 1 | Tianshi | | 84805 | 260. | 7.96969 | 27 | | 2290.22 | 12 | | | |
| Field 2 | Tianshi, by side of Tsung-Cheng | | 86742 | 265.868 | 4.56733 | 4 | | 50.2655 | 0 | 3 | 0 | 6 |

(*continued on next page*)

**Table 5** (*continued*).

| Field ID | in asterism | circNo | HIP | RA2000 | DE2000 | radius /° | comment | area/ °2 | ♯pairs | CVs | SNRs | PSRs |
|---|---|---|---|---|---|---|---|---|---|---|---|---|
| Field 1 | Wei | | 84638 | 259.55 | −39.6908 | 5 | | 78.5398 | 2 | – | – | – |
| Field 2 | Wei | | 84150 | 258.068 | −39.5069 | 8 | | 201.062 | 2+9 | +14 | +17 | ~100 |
| Field 3 | Wei, betw. 2nd and 3rd star | | 82554 | 253.114 | −40.723 | 3 | north of zeta Sco = HIP 82729 | 78.5398 | | | | |
| | Wei, betw. 2nd and 3rd star | | 82514 | 252.968 | −38.0474 | 4 | alternatively | | | | | |
| | Wuche | 1 | 21010 | 67.5836 | 28.1319 | 7 | | | | | | |
| Field 1 | Wuche | 2 | 29949 | 94.5703 | 46.3604 | 3 | evening (alternatively) | 260.752 | 0 | - | – | – |
| | Wuche | 3 | 22545 | 72.7888 | 48.7407 | 5 | morning (alternatively) | | | | | |
| Field 2 | Wuche, north of | | 26569 | 84.7365 | 49.4162 | 5 | | 78.5398 | 0 | – | – | – |
| | Yi | 1 | 55795 | 171.475 | −13.7513 | 8 | | | | | | |
| Field 1 | Yi | 2 | 55636 | 170.965 | −21.264 | 8 | | 356.735 | 0 | 6 | 0 | 4 |
| | Yi | 3 | 53740 | 164.944 | −18.2988 | 4 | | | | | | |
| | Yi | 4 | 57283 | 176.191 | −18.3507 | 4 | | | | | | |
| Field 1 | Yingshi | 1 | 113829 | 345.771 | 20.9186 | 7 | | 307.876 | 0 | 9 | 0 | 9 |
| | Yingshi | 2 | 113881 | 345.944 | 28.0828 | 7 | | | | | | |
| Field 1 | Zhen | | 60486 | 186.009 | −19.5718 | 6.5 | | 132.732 | 0 | 3 | 0 | 1 |
| Field 1 | Zigong, …  …between Doushu and [Bei]Ji | | 58952 | 181.313 | 76.9057 | 6 | | 113.097 | 0 | 3 | 0 | 4 |
| | Zigong | 1 | 72607 | 222.676 | 74.1555 | 18 | | | | | | |
| Field 2 | Zigong | 2 | 37391 | 115.127 | 87.0201 | 18 | | 2165.63 | 2 | | 3 | 18 |
| | Zigong | 3 | 17959 | 57.5896 | 71.3323 | 11 | | | | | | |

We expect similar or even stronger variations of the peak magnitude in ancient naked-eye observations, biased towards the brightest events. Indeed, we found several cases of known novae in our historical search fields with relatively faint CV remnants: If some of them have flared up to naked-eye visibility some centuries ago, this would imply more cases of a rather large variability in peak magnitudes, e. g. also 6 mag in case of KT Eri (Hoffmann and Vogt, 2020c, tab. 9).

A related problem is the identification of V529 Ori with any of the suggested historical (modern) sightings. If Hevelius's nova in 1678 refers to V529 Ori, its peak brightness in the 17th century had been around 6 mag. As it now has ~19 mag, this leads to an unusually high amplitude of 13 mag exceeding the known range of $N_r$ amplitudes by ≥1 mag.

This immediately leads to another interesting question: Is the hibernation scenario (Shara et al., 1986) important in the context of recurrent novae considered at millennia time scales?

It is known that a classical nova eruption has strong effects on the accretion disk and the mass transfer within the system. It could also cause an orbital widening of the binary in a way that does not allow any mass transfer which would prevent the system from further nova eruptions. However, a subsequent loss of energy in the system, e. g. by the emission of gravitational waves, could kindle the mass transfer after a while. The pause of untouched Roche volumes is called hibernation, expected lasting several millennia, but there could be further mass transfer variations causing epochs between moderate and extreme quiescence. For instance, there is evidence of incipient dwarf nova activity a few years before a nova eruption (Mróz et al., 2016), as well as several decades after return to quiescent state (Honeycutt et al., 1998; Vogt et al., 2018).

In our sample of historical data, we found some small asterisms that are mentioned only a few times but then not any more for centuries. For instance, the small asterism Zhen (IAU-constellation Corvus) covers only $\sim 133°^2$ and is in an area with dark celestial background. There had been two transients in the 3rd century, 247 and 275, but it is not mentioned afterwards. Ho (1962, p. 155, 157) considers them both comets but Xu et al. (2000, p. 132) enlists the second of them as potential stellar transients (while the first one is reported to have a size of 1° which could be a hint on a tail or on rays due to brightness; cf. Protte and Hoffmann (2020, A1)). If there was a nova that erupted two times within 28 years (just like the period of some known $N_r$) but never before or afterwards, this could point to such a scenario of pausing eruptions on the time scale of millennia.

For similar reasons, the case of Beidou (Big Dipper in UMa) is interesting where two reports are preserved from 158 and 305 CE ($T$ = 147 years) and again two reports from 1123 and 1221 ($T$ = 98 years). In this case, the search field is remarkably bigger ($263°^2$) and it is well possible that all four cases refer to different sources. Still, if they have been caused by the same object, it could be a hint on the object's longterm evolution and/ or a strong variability of peak brightnesses.

In the Chinese asterism of Taiwei, we found three different possible sequences with periods between 71 and 258 years. All of them are present only from the first to the seventh century of the common era. What happened before and afterwards? Did the eruption activity end or did a decreasing amplitude prevent naked-eye visibility?

At the current state of the art, it is impossible to answer these questions but this study shows the possibilities that are opened by this method: A significant extension of the temporal baseline of observations could reveal new and surprising insights in astrophysical processes which are impossible to achieve without systematic studies as those presented in our series of recent papers.

## 6. Data for the definition of the search fields

First, we append the information for deriving the search fields:

Each search field is addressed in the text with the name of the asterism and the number of the field (like the surname and given name of a person), e. g. 'Dongbi Field 2'. Thus, we had to define the Field *n* from reading the text and, then, cover this field by circles as we reasoned in Hoffmann et al. (2020).

In Table 5 our search 'Fields' are characterized by coordinates: The 'Fields' are in principle of random shape; they could be polygons, triangles, rectangles or whatever. However, for our catalogue search, we argued and decided in Hoffmann et al. (2020) to characterize them by covering them with circles.

## Data availability

The data underlying this article are available in the article and in its online supplementary material.

## Acknowledgements

This research has made use of 'Aladin sky atlas' developed at CDS, Strasbourg Observatory, France (Bonnarel et al., 2000; Boch and Fernique, 2014). Thankfully we made use of the Variable Star indeX VSX (Watson et al., 2006) of the American Association of Variable star Observers (AAVSO), the General Catalogue of Variable Stars (GCVS), the AAVSO Light Curve Generators https://www.aavso.org/LCGv2/, Light curve from CRTS/CSS (Drake et al., 2009), ASAS-SN Sky Patrol light curve (Shappee et al., 2014), and we acknowledge ESA Gaia, DPAC and the Photometric Science Alerts Team (http://gsaweb.ast.cam.ac.uk/alerts).

Furthermore, we thank the CDS Strasbourg for the VizieR and SIMBAD data base (Wenger et al., 2000) and happily used the Catalina Sky Survey (Drake et al., 2012) and the ATNF Pulsar Catalogue (Manchester et al., 2005) as well as Stellarium http://stellarium.org. The Curve Fitter https://statpages.info/nonlin.html was useful for the linear model fits although all other computations and selections were performed with own routines.

We thank Dieter B. Herrmann (Archenhold-Sternwarte Berlin) for his encouragement and advise.

S.H. acknowledges financial support from the Friedrich Schiller University and the Planetarium Jena, Germany. N.V. acknowledges financial support from Centro de Astrofísica, Universidad de Valparaíso, Chile.

## References

Ashbrook, J., 1963. Sky Telesc. 26 (N2), 80.
Boch, T., Fernique, P., 2014. Aladin lite: Embed your sky in the browser. p. 277.
Bode, M.F., Evans, A. (Eds.), 1989, 2008. Classical Novae. Cambridge University Press,
Bond, H.E., Miszalski, B., 2018. Publ. Astron. Soc. Pac. 130 (991), 094201.
Bonnarel, F., Fernique, P., Bienaymé, O., Egret, D., Genova, F., Louys, M., Ochsenbein, F., Wenger, M., Bartlett, J., 2000. Astron. Astrophys. Suppl. Ser. 143, 33–40.
Clark, D., Stephenson, F., 1977. The Historical Supernovae. Pergamon, Oxford.
Darnley, M.J., 2019. Accrete, accrete, accrete... Bang! (and repeat): The remarkable recurrent novae. arXiv e-prints. arXiv:1912.13209.
della Valle, M., 1991. Astron. Astrophys. 252, L9.
Drake, A.J., Djorgovski, S.G., Mahabal, A., Beshore, E., Larson, S., Graham, M.J., Williams, R., Christensen, E., Catelan, M., Boattini, A., Gibbs, A., Hill, R., Kowalski, R., 2009. Astrophys. J. 696 (1), 870–884.
Drake, A.J., Djorgovski, S.G., Mahabal, A., Prieto, J.L., Beshore, E., Graham, M.J., Catalan, M., Larson, S., Christensen, E., Donalek, C., Williams, R., 2012. In: Griffin, E., Hanisch, R., Seaman, R. (Eds.), New Horizons in Time Domain Astronomy. In: IAU Symposium, 285, pp. 306–308. http://dx.doi.org/10.1017/S1743921312000889, arXiv:1111.2566.
Duerbeck, H.W., 1987. Astrophys. Space Sci. 131 (1–2), 461–466.
Fekel, F.C., Hinkle, K.H., Joyce, R.R., Wood, P.R., Lebzelter, T., 2007. Astron. J. 133 (1), 17–25.
Ferrand, G., Safi-Harb, S., 2012. Adv. Space Res. 49 (9), 1313–1319.
Formiggini, L., Leibowitz, E.M., 1994. Astron. Astrophys. 292, 534–542.
Grasshoff, G., 1990. The History of Ptolemy's Star Catalogue. Springer, New York,
Griffith, D., Fabian, D., Sion, E.M., 1995. Publ. Astron. Soc. Pac. 107, 856.
Ho, P.Y., 1962. Vistas in Astronomy 5.
Hoffmann, S.M., 2017. Hipparchs Himmelsglobus. Springer.
Hoffmann, S.M., 2019. Mon. Not. R. Astron. Soc. 490 (3), 4194–4210.

Hoffmann, S.M., Vogt, N., 2020a. Mon. Not. R. Astron. Soc. 497 (2), 1419–1433.
Hoffmann, S.M., Vogt, N., 2020b. Mon. Not. R. Astron. Soc. 494 (4), 5775–5786.
Hoffmann, S.M., Vogt, N., 2020c. Mon. Not. R. Astron. Soc. 496 (4), 4488–4506.
Hoffmann, S., Vogt, N., 2021. In: Wolfschmidt, G., Hoffmann, S.M. (Eds.), Applied and Computational Historical Astronomy. In: Nuncius Hamburgensis, vol. 55, tredition, Hamburg,
Hoffmann, S.M., Vogt, N., Protte, P., 2020. Astron. Nachr. 341 (1), 79–98.
Honeycutt, R.K., Robertson, J.W., Turner, G.W., 1998. Astron. J. 115 (6), 2527–2538.
Hsi, T.-T., 1957. Smithonian Contrib. Astrophys. 2.
Latusseck, A., Hoffmann, S.M., 2017. Ein nützliches Unternehmen. Albireo, Cologne,
Leibowitz, E.M., Formiggini, L., 2013. Astron. J. 146 (5), 117.
Limoges, M.M., Bergeron, P., Lépine, S., 2015. Astrophys. J.s 219 (2), 19.
López-Gil, N., Rucker, F.J., Stark, L.R., Badar, M., Borgovan, T., Burke, S., Kruger, P.B., 2007. Vis. Res. 47 (6), 755–765, URL http://www.sciencedirect.com/science/article/pii/S0042698906003488.
Manchester, R.N., Hobbs, G.B., Teoh, A., Hobbs, M., 2005. Astron. J. 129 (4), 1993–2006.
Mróz, P., Udalski, A., Pietrukowicz, P., Szymański, M.K., Soszyński, I., Wyrzykowski, Ł., Poleski, R., Kozłowski, S., Skowron, J., Ulaczyk, K., Skowron, D., Pawlak, M., 2016. Nature 537 (7622), 649–651.
Munari, U., 2019. The symbiotic stars. arXiv e-prints, arXiv:1909.01389.
Murset, U., Nussbaumer, H., 1994. Astron. Astrophys. 282, 586–604.
Packer, D.E., 1894. J. Br. Astron. Assoc. 4, 364–368.
Pliny, Bostock, J., 1855. The Natural History. Taylor and Francis,
Prialnik, D., Kovetz, A., 2005. In: Burderi, L., Antonelli, L.A., D'Antona, F., di Salvo, T., Israel, G.L., Piersanti, L., Tornambè, A., Straniero, O. (Eds.), Interacting Binaries: Accretion, Evolution, and Outcomes. In: American Institute of Physics Conference Series, vol. 797, pp. 319–330. http://dx.doi.org/10.1063/1.2130250.
Protte, P., Hoffmann, S.M., 2020. Astron. Nachr. 341 (8), 827–840.
Ringwald, F.A., Naylor, T., 1997. Astron. Astrophys. 326, 629–631.
Robertson, J.W., Honeycutt, R.K., Hillwig, T., Jurcevic, J.S., Henden, A.A., 2000. Astron. J. 119 (3), 1365–1374.
Schaefer, B.E., 2010. Astrophys. J.s 187 (2), 275–373.
Shackleton, W., Fowler, A., 1894. J. Br. Astron. Assoc. 4, 215.

Shappee, B.J., Prieto, J.L., Grupe, D., Kochanek, C.S., Stanek, K.Z., De Rosa, G., Mathur, S., Zu, Y., Peterson, B.M., Pogge, R.W., Komossa, S., Im, M., Jencson, J., Holoien, T.W.-S., Basu, U., Beacom, J.F., Szczygieł, D.M., Brimacombe, J., Adams, S., Campillay, A., Choi, C., Contreras, C., Dietrich, M., Dubberley, M., Elphick, M., Foale, S., Giustini, M., Gonzalez, C., Hawkins, E., Howell, D.A., Hsiao, E.Y., Koss, M., Leighly, K.M., Morrell, N., Mudd, D., Mullins, D., Nugent, J.M., Parrent, J., Phillips, M.M., Pojmanski, G., Rosing, W., Ross, R., Sand, D., Terndrup, D.M., Valenti, S., Walker, Z., Yoon, Y., 2014. Astrophys. J. 788 (1), 48.
Shara, M.M., Drissen, L., Martin, T., Alarie, A., Stephenson, F.R., 2017a. Mon. Not. R. Astron. Soc. 465 (1), 739–745.
Shara, M.M., Iłkiewicz, K., Mikołajewska, J., Pagnotta, A., Bode, M.F., Crause, L.A., Drozd, K., Faherty, J., Fuentes-Morales, I., Grindlay, J.E., Moffat, A.F.J., Pretorius, M.L., Schmidtobreick, L., Stephenson, F.R., Tappert, C., Zurek, D., 2017b. Nature 548 (7669), 558–560.
Shara, M.M., Livio, M., Moffat, A.F.J., Orio, M., 1986. Astrophys. J. 311, 163.
Skopal, A., Chochol, D., Pribulla, T., Vanko, M., 2000. Inf. Bull. Var. Stars 5005, 1.
Skopal, A., Shugarov, S.Y., Munari, U., Masetti, N., Marchesini, E., Komžík, R., Kundra, E., Shagatova, N., Tarasova, T., Buil, C., Boussin, C., Shenavrin, V., Hambsch, F.J., Dallaporta, S., Frigo, A., Garde, O., Zubareva, A., Dubovský, P., Kroll, P., 2020. Astron. Astrophys. 636, A77.
Vogt, N., 1990. Astrophys. J. 356, 609.
Vogt, N., Hoffmann, S.M., Tappert, C., 2019. Astron. Nachr. 752, 340.
Vogt, N., Tappert, C., Puebla, E.C., Fuentes-Morales, I., Ederoclite, A., Schmidtobreick, L., 2018. Mon. Not. R. Astron. Soc. 478 (4), 5427–5435.
Warner, B., 1995. Cataclysmic Variable Stars. In: Cambridge Astrophysical Series, Cambridge University Press.
Watson, C.L., Henden, A.A., Price, A., 2006. Soc. Astron. Sci. Annu. Symp. 25, 47.
Wenger, M., Ochsenbein, F., Egret, D., Dubois, P., Bonnarel, F., Borde, S., Genova, F., Jasniewicz, G., Laloë, S., Lesteven, S., Monier, R., 2000. Astron. Astrophys. Suppl. Ser. 143, 9–22.
Woudt, P.A., Ribeiro, V.A.R.M., 2014. In: Woudt, P.A., Ribeiro, V.A.R.M. (Eds.), Stellar Novae: Past and Future Decades. In: Astronomical Society of the Pacific Conference Series, vol. 490.
Xu, Z., Pankenier, D.W., Jiang, Y., 2000. East Asian Archaeoastronomy. Gordon and Breach Science Publishers,
Yaron, O., Prialnik, D., Shara, M.M., Kovetz, A., 2005. Astrophys. J. 623 (1), 398–410.